\documentstyle[preprint,revtex,eqsecnum]{aps}
\begin{document}
\preprint{CALT 68-1848}
\draft
\def\intof{\int_{t_0}^{t_f}}
\def\intot{\int_{t_0}^t}
\def\Ss{S_{\rm sys}}
\def\Sr{S_{\rm res}}
\def\Si{S_{\rm int}}
\def\dLdX{{{\partial L}\over{\partial X}}}
\def\dLdXd{{{\partial L}\over{\partial {\dot X}}}}
\def\dLdx{{{\partial L}\over{\partial x}}}
\def\dLdxd{{{\partial L}\over{\partial {\dot x}}}}
\def\dVkdr{{{\partial V_k}\over{\partial Q}}}
\def\ddVkddr{{{\partial^2 V_k}\over{\partial Q^2}}}
\def\dVkdx{{{\partial V_k}\over{\partial x}}}
\def\xp{x^\prime}
\def\Qp{Q^\prime}
\def\delQ{\delta Q}
\def\delQp{\delta Q^\prime}
\def\delQf{\delta(Q^\prime(t_f) - Q(t_f))}
\def\e{{\rm e}}
\def\im{{\rm Im}}
\def\bra#1{\langle #1 |}
\def\ket#1{| #1 \rangle}
\def\braket#1#2{\langle #1 | #2 \rangle}
\def\ensemble#1{\langle #1 \rangle}

\begin{title}
Quasiclassical Equations of Motion \\
for Nonlinear Brownian Systems
\end{title}

\bigskip
\bigskip

\author{Todd A. Brun}
\begin{instit}
California Institute of Technology,
Pasadena, CA  91125
\end{instit}

\bigskip
\bigskip

\nonum\section{\bf Abstract}
\begin{abstract}
Following the formalism of Gell-Mann and Hartle, phenomenological equations
of motion are derived from the decoherence functional of quantum mechanics,
using a path-integral description.
This is done explicitly for the case of a system interacting with a ``bath''
of harmonic oscillators whose individual motions are neglected.
The results are compared to the equations
derived from the purely classical theory.
The case of linear interactions is treated exactly, and nonlinear interactions
are also compared, using classical and quantum perturbation theory.
\end{abstract}

\vfill\eject

\section{\bf Introduction}
\subsection{Decoherence and the quasiclassical limit}

Two of the most puzzling aspects of the quantum theory have, until recently,
remained unclear: the proper interpretation of quantum probabilities, and
the mechanism by which deterministic classical ``laws'' can arise from a
probabilistic underlying theory.  The idea of wave-function collapse,
while providing a useful approximate description of most experimental
situations, begs the question of why a system which otherwise undergoes
purely unitary evolution should suddenly and dramatically be collapsed upon
measurement by a scientist.  The procedure is highly asymmetric, instantaneous,
and irreversible, and moreover requires the existence of a ``classical''
measuring device outside the system being measured.  When one considers a
closed system the idea of wave-functions
collapsing becomes highly ambiguous.  There is nothing outside the system
to collapse it.  The quintessential example of this, of course, is the
universe itself.  Clearly, if the fundamental laws of the universe are
quantum mechanical, there can be no separate ``classical domain'' to
explain our observations.
Since the classical realm is itself, presumably, merely
a limit of the underlying quantum reality, the probabilities must arise
directly from the quantum theory itself, without recourse to the deus
ex machina of the measurement device.  And somehow, the various potential
futures of the universe must collapse themselves onto the one possibility
which we observe.
Quantum cosmology requires a solid formalism for
the treatment of closed systems, and work in this field should have that
as its goal.

The recent work on the decoherence functional formulation avoids the
problems of earlier approaches \cite{Griffiths,GMHart0,GMHart1,Omnes}.
Physics is described in terms of exhaustive sets of possible
histories, coarse-grained, with the restriction that these histories must
be decoherent.  That is, it must be possible to assign probabilities to these
histories such that they obey the classical laws of probability, with no
interference.

Gell-Mann and Hartle have argued that it is possible, in a highly
coarse-grained system, to define the classical equation of motion directly
from the decoherence functional itself \cite{GMHart2}.
I will, in this paper, attempt
to show that this definition gives the exact classical results, at least
for the case of systems interacting with baths of oscillators;
and further, that these systems are
decoherent in the classical limit.  Quantum effects enter as a random,
fluctuating force from the effects of neglected degrees of freedom,
even in cases where the classical noise would ordinarily be zero.
Fluctuations, dissipation, and decoherence turn out to be intimately
interlinked.

The linear case has been treated before by a number of people, both
classically and quantum mechanically, though not in precisely this
same way \cite{FeynVern,Zwanzig,CaldLegg}.
The correspondence of this quantum system to the classical
Langevin equation is thus nothing new.  The decoherence of similar
systems has also been examined, using a somewhat different definition
of decoherence which for these models generally corresponds to my
definition \cite{UnZur}.  However, to my knowledge, no one has considered the
classical correspondence of these sorts of nonlinear systems, nor
the relationship between dissipation, noise, and decoherence in these
more general cases.  Thus, the results herein are of interest in
demonstrating that it is possible to define classical equations of
motion directly from the quantum theory in a broad range of systems.

\subsection{Path-integral description of the decoherence functional}

We will not, for the purposes of this model, be using the decoherence
functional in its most general form.  Instead, we will consider only one
type of history.  Suppose that our system is completely described by a
set of generalized coordinates $q^\beta$ (collectively referred to as $q$).
The most fine-grained
possible family of histories would be just the set of all possible
paths $q(t)$.  We can coarse-grain this by dividing the range
of the $q^\beta$ into an exhaustive set of intervals $\Delta^i_{\alpha_i}(t_i)$
at a sequence of times $t_1,t_2,\ldots$, where the $\alpha_i$ are an index
labeling the intervals.  We can then specify one particular history by
which interval was passed through at each time,
labeling it by the sequence of indices
$\alpha_1,\alpha_2,\ldots$, which I will
generally abbreviate as $\alpha$.  Such a history includes all possible
paths which pass through the given set of intervals at the given times.

The decoherence functional is a functional on pairs of histories.  The
value of this functional on a pair of histories
$\alpha$ and $\alpha^\prime$ is given by

\begin{equation}
D(\alpha^\prime,\alpha) = \int_{\alpha^\prime} \delta q^\prime
\int_{\alpha} \delta q \ \delta(q^{\prime}_f - q_f) \exp
\biggl\{ {i({S[q^\prime(t)] - S[q(t)]})/\hbar} \biggr\}
\rho(q_0^\prime,q_0).
\end{equation}

Here we are integrating over all paths which pass through the specified
sequence of intervals.  The functional $S[q(t)]$ is the fundamental action.
If ${\rm Re} D(\alpha^\prime,\alpha) = 0$ for $\alpha^\prime \ne \alpha$,
then the system is said to be {\it decoherent}, and obeys classical laws
of probability.  The diagonal elements $D(\alpha,\alpha)$ are the
probabilities of each history $\alpha$.

The simplest form of this type of history is that where the intervals are
completely fine-grained in certain variables, and completely coarse-grained
in others.  We divide the coordinates $q^\beta$ into two groups, $x^\beta$
(henceforth known as the {\it system} coordinates),
referred to collectively as $x$, and $Q^k$ (henceforth
known as the {\it reservoir} coordinates),
referred to collectively as $Q$.  Our histories will then be
complete trajectories $x^\beta(t)$ for the system coordinates, while the
reservoir coordinates will be neglected completely.

It is then convenient to break the fundamental action of the system into
several parts:
\begin{equation}
S[q(t)] = S_{\rm sys}[x(t)] + S_{\rm res}[Q(t)]
- \int_{t_0}^{t_f} V(x(t),Q(t))\ dt,
\end{equation}
where $S_{\rm sys}[x(t)]$ is the action of the system, $S_{\rm res}[Q(t)]$
is the action of the reservoir, and there is an interaction potential
$V(x,Q)$ between them.  The decoherence functional is then

\vfill\eject
\begin{eqnarray}
D[\xp(t),x(t)] =&& \exp \biggl\{ i(S_{\rm sys}[\xp(t)] -
  S_{\rm sys}[x(t)])/\hbar \biggr\} \int \delQp \int \delQ \
  \delQf \nonumber\\
&&\times\exp \biggl\{ i(S_{\rm res}[\Qp(t)] - S_{\rm res}[Q(t)] \nonumber\\
&&\ \ - \int_{t_0}^{t_f} (V(\xp(t),\Qp(t)) -
  V(x(t),Q(t)))\ dt)/\hbar\biggr\} \nonumber\\
&&\times\rho(\xp_0,\Qp_0; x_0,Q_0).
\end{eqnarray}

\section{\bf Linear case}

The case of a system interacting linearly with a reservoir is a famous one, and
has been treated by a number of people; quantum mechanically by Feynman and
Vernon, and Caldeira and Leggett, classically by Zwanzig.  For convenience,
it is customary to make a number of simplifying assumptions:
\smallskip

1.  The reservoir variables $Q^k$ are harmonic oscillators, i.e.,
\begin{equation}
\Sr[Q(t)] = \sum_k \int_{t_0}^{t_f} {m\over 2}\bigl({\dot {Q^k}}^2 -
\omega_k^2 {Q^k}^2\bigr)\ dt .
\end{equation}

2.  The initial density matrix factors:
\begin{equation}
\rho(\xp_0,\Qp_0;x_0,Q_0)
= \chi(\xp_0,x_0)\phi(\Qp_0,Q_0).
\end{equation}

A similar assumption classically is to assume that the initial
probability distribution of the reservoir coordinates is independent
of the initial state of the system coordinates.

3.  The interaction $V(x,Q)$ is bilinear:
\begin{equation}
V(x,Q) = - \sum_k \gamma_k x Q^k.
\end{equation}

I will generally assume that $x$ is a single variable; multivariable systems
are a trivial generalization, where the $\gamma_k$ become matrices.
\smallskip

We will relax these assumptions to a certain degree later on, but for now
let us consider this case.  The classical case is exactly solvable.  In
this, the equation of motion for the reservoir variable $Q^k$ is
\begin{equation}
{{d^2 Q^k}\over{dt^2}}(t) = - \omega_k^2 Q^k(t) + (\gamma_k/m) x(t).
\end{equation}
This has a solution
\begin{equation}
Q^k(t) = Q^k(t_0) \cos(\omega_k(t-t_0))
  + {{{\dot {Q^k}}(t_0)}\over{\omega_k}} \sin(\omega_k(t-t_0))
  + {{\gamma_k}\over{m\omega_k}}\intot \sin[\omega_k(t-s)] x(s)\ ds.
\end{equation}
We can then use this in the equation of motion for $x$:
\begin{eqnarray}
{d\over{dt}}\biggl(\dLdxd \biggr)(t) &&= \biggl(\dLdx \biggr)(t)
  + \sum_k \gamma_k Q^k(t) \nonumber\\
&&= \biggl(\dLdx \biggr)(t) + F(t) + \sum_k {{\gamma_k^2}\over{m\omega_k}}
  \intot \sin[\omega_k(t-s)] x(s)\ ds, \label{classeqn}
\end{eqnarray}
where $F(t)$ is the {\it Langevin force}.  In this case, it is clearly
\begin{equation}
F(t) = \sum_k \gamma_k \biggl( Q_k(t_0) \cos(\omega_k(t-t_0))
  + {{{\dot {Q^k}}(t_0)}\over{\omega_k}} \sin(\omega_k(t-t0))
\biggr).\label{stoch}
\end{equation}

If we assume that the $Q$ have a thermal probability distribution initially,
\begin{equation}
P(Q) = \prod_k {{m\omega_k}\over{2\pi kT}}\exp\left(-{m\over{2kT}}
({\dot {Q^k}}^2 + \omega_k^2 {Q^k}^2)\right),
\end{equation}
which, when averaged over an ensemble, gives
\begin{mathletters}
\begin{eqnarray}
\ensemble{Q^k} = 0,\ \  \ensemble{{Q^k}^2} =
  {{kT}\over{m\omega_k^2}},\ \ldots, \\
\ensemble{{\dot {Q^k}}} = 0,\ \  \ensemble{{\dot {Q^k}}^2} =
  {{kT}\over{m}},\ \ldots,
\end{eqnarray}
\end{mathletters}
then
\begin{mathletters}
\begin{eqnarray}
\ensemble{F(t)} =&& 0,\\
\ensemble{F(t)F(s)} =&& \sum_k \gamma_k^2 \biggl( {{kT}\over{m\omega_k^2}}
  \biggr) \cos(\omega_k(t-s)). \label{correl}
\end{eqnarray}
\end{mathletters}

Let us compare this to the quantum results.
It is interesting to first consider the
system in isolation from the reservoir.  In this case we would have
\begin{equation}
D[\xp(t),x(t)] = \exp\biggl\{i(S_{\rm sys}[\xp(t)] -
S_{\rm sys}[x(t)])/\hbar)\biggr\}
\chi(\xp_0,x_0).
\end{equation}
If the action $\Ss[x(t)]$ has the usual form
\begin{equation}
\Ss[x(t)] = \int_{t_0}^{t_f} L(x(t),{\dot x}(t))\ dt,
\end{equation}
then we can change variables
\begin{mathletters}
\begin{eqnarray}
X(t) &&= {1\over2}[\xp(t) + x(t)], \\
\xi(t) &&= \xp(t) - x(t),
\end{eqnarray}
\end{mathletters}
and expand the phase in terms of $\xi$:
\begin{eqnarray}
\Ss[\xp(t)] - \Ss[x(t)] &&=
\intof \dLdX\bigl(X(t),{\dot X}(t)\bigr)\xi(t) +
\dLdXd\bigl(X(t),{\dot X}(t)\bigr){\dot \xi}(t) + {\rm O}(\xi^3)\ dt\nonumber\\
&&= \intof \biggl(-{d\over{dt}}\dLdXd\bigl(X(t),{\dot X}(t)\bigr) +
\dLdX\bigl(X(t),{\dot X}(t)\bigr)\biggr)\xi(t)\ dt \nonumber\\
&&\ \ -\dLdXd\bigl(X_0,{\dot X}_0\bigr)\xi(t_0)
  + {\rm O}(\xi^3).
\end{eqnarray}
So the Euler-Lagrange equation of motion appears in the phase of the
decoherence functional!

One should not put too much weight on this occurrence.  This system is not
decoherent; substantial interference can still occur between different
possible trajectories.  There is no particular reason to expect $\xi(t)$
to be small, so it is not correct to neglect higher-order terms.  This
system, on its own, is still essentially quantum-mechanical.  It is not
even quasiclassical.

This still leaves the effects of the reservoir variables and interaction
unaccounted for.  Let us turn, then, to this portion of the decoherence
functional.
\begin{eqnarray}
F[\xp(t),x(t)] =&& \int\delQp\int\delQ\delQf
\exp\biggl\{ i\biggl(\Sr[\Qp(t)] -
\Sr[Q(t)] \nonumber\\
&&\ \ - \intof V(\xp(t),\Qp(t)) -
V(x(t),Q(t))\ dt\biggr)/\hbar\biggr\}
\chi(\Qp_0,Q_0) \nonumber\\
=&& \exp\biggl\{i W[\xp(t),x(t)]/\hbar \biggr\}.
\end{eqnarray}
$F[\xp(t),x(t)]$ is termed the {\it influence functional} by Feynman
and Vernon, and $W[\xp(t),x(t)]$ is the {\it influence phase} \cite{FeynVern}.
In our
simplified model, this is not difficult to evaluate exactly.  It is generally
assumed that the initial density matrix is in a thermal state.
We quote the results
of Feynman and Vernon:
\begin{equation}
W[\xp(t),x(t)] = {1\over2}\intof dt \intot ds\
[\xp(t) - x(t)]\biggl(k(t-s)\xp(s) + \*k(t-s)x(s)\biggr),
\end{equation}
where the real and imaginary parts of $k(t-s)$ are
\begin{mathletters}
\begin{eqnarray}
k_R(t-s) &&= \sum_k {{\gamma_k^2}\over{m\omega_k}}
 \sin(\omega_k (t-s)),\\
k_I(t-s) &&= \sum_k {{\gamma_k^2}\over{m\omega_k}} \coth(\hbar\omega_k/kT)
\cos(\omega_k (t-s)).
\end{eqnarray}
\end{mathletters}
Changing to our variables $X$ and $\xi$, we see that
\begin{eqnarray}
W[X(t),\xi(t)] =&& \sum_k {{\gamma_k^2}\over{2m\omega_k}}
\int_{t_0}^{t_f} dt \int_{t_0}^{t_f}\ ds
\biggl\{2\xi(t)X(s)\sin(\omega_k(t-s)) \nonumber\\
&&\ \ + i\xi(t)\xi(s)\coth\left({{\hbar\omega_k}\over{kT}}\right)
\cos(\omega_k(t-s)) \biggr\}.
\end{eqnarray}
Thus, we have a real term which is proportional to $\xi(t)$ and an
imaginary term which is proportional to $\xi(t)\xi(s)$.  The imaginary
term is a double integral over a symmetric kernel whose eigenvalues are
strictly
non-negative; thus, for large $\xi$ the decoherence functional will be
diminished by a decaying exponential
\[
\exp\biggl[-\intof dt\intot ds\ \xi(t)\xi(s)\cos(\omega_k(t-s))\biggr].
\]
Since $\xi$ essentially
measures how far you are from the diagonal of the decoherence functional,
the off-diagonal terms tend to vanish and the system becomes decoherent.

Furthermore, since large $\xi$ is suppressed, it now makes sense to discard
terms of $O(\xi^3)$.  Thus we can now say
\begin{eqnarray}
\Ss[\xp(t)] - \Ss[x(t)] + W[\xp(t),x(t)] =&&
 {i\over4}\intof dt \intof ds\
\xi(t)k_I(t-s)\xi(s) \nonumber\\
&&\ \ + \intof dt\ \xi(t)e(t)
+ O(\xi^3),
\end{eqnarray}
where
\begin{equation}
e(t) = -{d\over dt}\biggl(\dLdXd(t)\biggr) + \dLdX(t) -
\sum_k {{\gamma_k^2}\over{m\omega_k}}
\int_{t_0}^{t} ds \ X(s)\sin(\omega_k(t-s)).
\end{equation}
If we compare this to (\ref{classeqn}), we see that
\[
e(t) = 0
\]
is identical to the ensemble-averaged classical equation of motion.
Note that the bath of harmonic oscillators acts as a retarded force on
the system.  In the limit as we go to a continuum of oscillator frequencies
with a high cut-off, this retarded force becomes a dissipative term, i.e.,
a frictional force.  In this limit, Caldeira and Leggett show that
for a Debye distribution of oscillator frequencies, the
influence phase becomes \cite{CaldLegg}
\begin{equation}
W[X,\xi] =  \intof \biggl( - 2 \Gamma {\dot X}\xi(t)
  + {{i kT}\over\hbar} \Gamma\xi^2(t) \biggr) dt.
\end{equation}
where $\Gamma$ is the usual classical coefficient of friction, defined
in terms of $\gamma$ and the cutoff frequency $\Omega$.  See
\cite{Zwanzig,CaldLegg} for details.

We have seen that the real term of $W[X,\xi]$ corresponds to the classical
retarded or (in the limit) dissipative force.
The imaginary term also has a classical
analog.  In the classical case, there is a random stochastic force $F(t)$
given by (\ref{stoch}),
which ensemble-averages to zero $\ensemble{F(t)} = 0$.
As we see in (\ref{correl}), however, the two-time correlation function of this
force does not vanish.
As $\hbar \rightarrow 0$, we get $\coth(\hbar\omega/2kT) \rightarrow
2kT/\hbar\omega$.  So the imaginary part of $W[X,\xi]$ has the form
\begin{equation}
{\rm Im}\ W[X(t),\xi(t)] = \intof dt \intot ds\
\ensemble{F(t)F(s)}\xi(t)\xi(s).
\end{equation}

Here we observe the subtle linkage between noise, dissipation, and decoherence.
In interacting with the many degrees of freedom of the reservoir, the system
loses energy.  It also is subject to random jostlings from the reservoir
oscillators.  But one last, purely quantum-mechanical effect is that the
state of the system is continually being ``measured,'' and thus the various
possible trajectories tend to decohere, at least on a scale large compared
to $\hbar$.  Later we will see that even in situations where the classical
noise vanishes, there is still quantum-mechanical noise.  This arises
essentially from the zero-point energy of the reservoir oscillators.

We can straightforwardly generalize to the case where the potential is
nonlinear in $x$, but still linear in $Q$.  Suppose that
\begin{equation}
V(x,Q) = - \sum_k a_k(x) Q^k.
\end{equation}
Here the influence phase is
\begin{eqnarray}
W[\xp(t),x(t)] =&&  \sum_k \intof dt \intot ds\ \biggl\{
(a_k(\xp(t)) - a_k(x(t))) \nonumber\\
&&\ \ \times(a_k(\xp(s))k_k(t-s) -
a_k(x(s))k_k^\ast(t-s)) \biggr\},
\end{eqnarray}
where
\begin{equation}
k_k(t-s) = {1\over{2m\omega_k}}\biggl[ \sin(\omega_k(t-s)) +
i \coth(\hbar\omega_k/2kT) \cos(\omega_k(t-s)) \biggr].
\end{equation}
We can again separate the real and imaginary parts, and change to variables
$X$ and $\xi$.  We then get, to $O(\xi^3)$,
\begin{eqnarray}
W[X(t),\xi(t)] = -{1\over{2m}}\sum_k \intof dt \intot ds\ \biggl\{
a_k^\prime(X(t)) a_k(X(s)) \xi(t) \sin(\omega_k(t-s)) \nonumber\\
- i\coth\left({{\hbar\omega_k}\over{2kT}}\right)
a_k^\prime(X(t)) a_k^\prime(X(s)) \xi(t)\xi(s)
\cos(\omega_k(t-s)) \biggr\}.
\end{eqnarray}

Again, we see that the real term has the same form as the classical retarded
force, which becomes dissipative in the limit of continuous frequencies and
high cutoff.  The imaginary term again corresponds to a double integral
over the two-time correlation function of the classical stochastic force.
It is strictly non-negative, and exponentially damps the decoherence
functional for large $\xi$.

\section{\bf Nonlinear examples}

The problem with potentials nonlinear in $Q$ is that the path integral
is no longer solvable in closed form.  Thus, it is difficult to be certain
that this correspondence with the classical equation of motion which holds
in the linear case is truly universal.  We can, however, consider weak
couplings, and solve for the equation of motion using perturbation theory.
We can then compare the classical perturbative equation to that derived
from the influence functional.

\subsection{Classical and Quantum perturbation theory}

Let us consider a system coupled to a bath of harmonic oscillators with
a potential of the form
\begin{equation}
V(x,Q) = - \epsilon \sum_k V_k(x,Q^k), \label{one}
\end{equation}
where $V_k(x,Q^k)$ can be nonlinear in $x$ and $Q^k$.
In general, such a problem
cannot be solved exactly.  However, if the coupling is weak
($\epsilon << 1$) then we can make a perturbation expansion, at least
for reasonably well-behaved potentials.

The total Lagrangian is
\begin{equation}
L_{\rm total}(x,{\dot x},Q,{\dot Q}) =
L(x,{\dot x}) + \sum_k {m\over2}\biggl({({\dot {Q^k}})}^2
- \omega_k^2 {(Q^k)}^2 \biggr) -
\epsilon V_k(x,Q^k).
\end{equation}
Let's suppose that the trajectory $x(t)$ is known.  Then the equation of
motion for the $k$th harmonic oscillator is
\begin{equation}
{{d^2Q^k}\over{dt^2}} = - \omega_k^2 Q^k +
{\epsilon\over m} \dVkdr(x(t),Q^k).
\end{equation}
If we then write $Q^k$ as an expansion
\begin{equation}
Q^k(t) = Q^k_0(t) + \epsilon Q^k_1(t) + \epsilon^2 Q^k_2(t) + \ldots
\end{equation}
and equate equal powers of $\epsilon$ we get a series of equations
\begin{mathletters}
\begin{equation}
{{dQ^k_0}\over{dt}} = - \omega_k^2 Q^k_0, \label{two}
\end{equation}
\begin{equation}
{{dQ^k_1}\over{dt}} = - \omega_k^2 Q^k_1
+ {1\over m}\dVkdr(x(t),Q^k_0(t)),
\end{equation}
\begin{equation}
{{dQ^k_2}\over{dt}} = - \omega_k^2 Q^k_2 +
 {1\over m}Q^k_1(t)\ddVkddr(x(t),Q^k_0(t)),
\end{equation}
\end{mathletters}
etc., where we've Taylor-expanded $V_k(x,Q^k_0 + \epsilon Q^k_1 + \cdots)$ in
powers of $\epsilon$.

Now we have equations for each $Q^k_i(t)$ in terms of the lower order
functions.  Notably, the lowest order equation is now a simple harmonic
oscillator, and we can solve for it easily in terms of the initial
conditions
\begin{equation}
Q^k_0(t) = A_k \cos(\omega_k t) + B_k \sin(\omega_k t), \label{three}
\end{equation}
where $A_k = Q^k |_{t_0}$ and $B_k = (1/\omega_k)(dQ^k/dt) |_{t_0}$.

The higher-order equations are driven oscillators.  We can
solve for them exactly, matching initial conditions:
\begin{mathletters}
\begin{equation}
Q^k_1(t) = {1\over{m\omega_k}} \intot \sin(\omega_k(t-s))
\dVkdr(x(s),Q^k_0(s))\ ds, \label{four}
\end{equation}
\begin{equation}
Q^k_2(t) = {1\over{m\omega_k}} \intot \sin(\omega_k(t-s))
\ddVkddr(x(s),Q^k_0(s)) Q^k_1(s)\ ds,
\end{equation}
\end{mathletters}
and so forth.

Having found the motion of the harmonic oscillators in terms of $x(t)$, we
can now turn around and find the equation of motion for $x$.  This is
\begin{equation}
{d\over{dt}}\dLdxd(t) = \dLdx(t) + \epsilon \sum_k
\dVkdx(x(t),Q^k(t)).
\end{equation}
$Q^k(t)$ is the expansion that we solved for above, and it will depend on
the earlier behavior of $x$, in general.  Note that causality is strictly
obeyed.  This classical causality follows as a result of
more fundamental quantum causality, as discussed by Gell-Mann and Hartle
\cite{GMHart2}.

We treat this same problem quantum-mechanically by trying to find the
influence functional $F[\xp(t),x(t)]$ as a perturbation expansion.
Assume that the reservoir starts in a definite initial state $\ket{a}$, with
wave function $\phi_a(Q)$.  Then
\begin{eqnarray}
F_a[\xp,x] =&& \int \delQf
\exp\left\{{i\over\hbar}
(\Sr[\Qp(t)] - \Sr[Q(t)])\right\} \nonumber\\
&&\ \ \times \biggl[1 + {{i\epsilon}\over\hbar}\intof \biggl(V(\xp(t),\Qp(t)) -
V(x(t),Q(t))\biggr)\ dt \nonumber\\
&&\ \ + \left({{i\epsilon}\over\hbar}\right)^2 \intof \intot
\biggl(V(\xp(t),\Qp(t))-V(x(t),Q(t))\biggr) \nonumber\\
&&\ \ \ \ \ \times\biggl(V(\xp(s),\Qp(s)) - V(x(s),Q(s))\biggr)\
ds\ dt + \cdots \biggr] \phi_a(\Qp(t_0))\phi^\ast_a(Q(t_0))
\ \delQp\ \delQ \nonumber\\
=&& 1 + {{i\epsilon}\over\hbar}\intof \biggl(V_{aa}(\xp(t))
- V_{aa}(x(t))\biggr)\ dt \nonumber\\
&&\ \ - \left({{\epsilon^2}\over{\hbar^2}}\right) \sum_b \intof \intot
\biggl(V_{ab}(\xp(t))V_{ba}(\xp(s))\e^{-i\omega_{ba}(t-s)} -
V_{ab}(x(t))V_{ba}(\xp(s))\e^{-i\omega_{ba}(t-s)} \nonumber\\
&&\ \ \ \ \ - V_{ba}(\xp(t))V_{ab}(x(s))\e^{i\omega_{ba}(t-s)} +
V_{ba}(x(t))V_{ab}(x(s))\e^{i\omega_{ba}(t-s)}\biggr)\ ds\ dt
+ \cdots \nonumber\\
=&& 1 + {{i\epsilon}\over\hbar}\intof V^\prime_{aa}(X(t))\xi(t)\ dt \nonumber\\
&&\ \ - \left({{\epsilon^2}\over{\hbar^2}}\right) \sum_b \intof \intot
\biggl(V^\prime_{ba}(X(t))V^\prime_{ab}(X(s))
\xi(t)\xi(s)\cos(\omega_{ba}(t-s)) \nonumber\\
&&\ \ \ \ \ - 2i V^\prime_{ba}(X(t))V_{ab}(X(s))
\xi(t)\sin(\omega_{ba}(t-s))\biggr)\ ds\ dt
+ O(\epsilon^3) + O(\xi^3). \label{five}
\end{eqnarray}
Here we've defined the functions
\begin{mathletters}
\begin{equation}
V_{aa}(x) = \bra{a}V(x,{\hat Q})\ket{a}
  = \int \phi_a(Q)\phi_a^\ast(Q)V(x,Q)\ dr,
\end{equation}
\begin{equation}
V_{ba}(x) = \bra{b}V(x,{\hat Q})\ket{a}
= \int \phi_a(Q)\phi_b^\ast(Q)V(x,Q)\ dr.
\end{equation}
\end{mathletters}

In our case, we assume that the reservoir is a collection of harmonic
oscillators initially in a thermal state.  In this case, the states $\ket{a}$
become the ordinary Fock states $\ket{n}$ and the influence functional is
\begin{equation}
F[\xp(t),x(t)] = \sum_n \rho_{nn} F_n[\xp(t),x(t)], \label{six}
\end{equation}
where
\begin{equation}
\rho_{nn} = \prod_k \biggl(1 - \exp(-\hbar\omega_k/kT)\biggr)
\exp(-n_k\hbar\omega_k/kT).
\end{equation}

\subsection{Polynomial potentials}

We will specifically consider a potential of the form (\ref{one}) where the
individual potentials are polynomials in $Q^k$.
We will see that it
is convenient to separate the even and odd terms:
\begin{equation}
V_k(x,Q^k) = \sum_{l=0}^N a_{kl}(x) {(Q^k)}^{2l+1}
+ \sum_{l=1}^N b_{kl}(x) {(Q^k)}^{2l}.
\end{equation}
The $a_{kl}(x)$ and $b_{kl}(x)$ are arbitrary functions of $x$, only
assuming that the potential as a whole remains relatively well-behaved,
integrable, etc.
For convenience, I will drop the index $k$ for the rest of this derivation.
It should be understood that the final result is to be summed over all
the oscillators,
\begin{equation}
W[\xp(t),x(t)] = \sum_k W_k[\xp(t),x(t)].
\end{equation}

{}From the equation (\ref{two}), we can write down the equations of motion for
a classical oscillator $Q(t) = Q_0(t) + \epsilon Q_1(t) + \ldots$.  We
then plug in the solutions (\ref{three}) and (\ref{four}) to get
\begin{mathletters}
\begin{equation}
Q_0(t) = A \cos(\omega t) + B \sin(\omega t) =
\beta \e^{i\omega t} + \beta^\ast \e^{-i\omega t},
\end{equation}
\begin{equation}
Q_1(t) = {1\over{m\omega}}\intot \sin[\omega(t-s)]
\left( \sum_{j=0}^N (2j+1) a_j(x(s)) Q_0^{2j}(s) +
\sum_{j=1}^N 2j b_j(x(s)) Q_0^{2j-1}(s) \right)\ ds, \label{seven}
\end{equation}
\end{mathletters}
etc., where
$\beta = (A - iB)/2$.
The equation of motion for $x$ is then
\begin{eqnarray}
{d\over{dt}}\left(\dLdxd \right) =&& \left(\dLdx \right) +
\epsilon\left( \sum_{j=0}^N a^\prime_j(x(t)) Q_0^{2j+1}(t) +
\sum_{j=1}^N b^\prime_j(x(t)) Q_0^{2j}(t) \right) \nonumber\\
&&\ \ + \epsilon^2\left( \sum_{j=0}^N (2j+1)a^\prime_j(x(t)) Q_0^{2j}(t)
+ \sum_{j=1}^N 2j b_j^\prime(x(t)) Q_0^{2j-1}(t) \right) Q_1(t)
+ O(\epsilon^3) \nonumber\\
=&&\left(\dLdx \right) + \epsilon\tau_1(t) + \epsilon^2\tau_2(t)
+ O(\epsilon^3) \label{eight}
\end{eqnarray}
We are interested in the ensemble-averaged equation.  We can make use of the
fact that
\begin{equation}
\ensemble{\beta^m {\beta^\ast}^n} =
  \delta_{mn} n!{\left({{kT}\over{2m\omega^2}}\right)}^n.
\end{equation}
So only the even terms contribute to the first-order component of the
equation (\ref{eight}).  $Q_0^{2j}(t)$ is readily found then with a binomial
expansion of
\begin{equation}
{(\beta\e^{i\omega t} + \beta^\ast\e^{-i\omega t})}^{2j} =
  \sum_{i=0}^{2j} {{2j}\choose{i}} {(\beta\e^{i\omega t})}^i
  {(\beta^\ast\e^{-i\omega t})}^{2j-i},
\end{equation}
\[
{{n}\choose{i}} = {{n!}\over{i!(n-i)!}},
\]
yielding
\begin{equation}
\ensemble{\tau_1(t)} = \sum_{j=1}^N {{2j!}\over{j!}}
  b_j^\prime(x(t)) {\left({kT\over{2m\omega^2}}\right)}^j. \label{nine}
\end{equation}

The second order component is more complicated.  Plugging expression
(\ref{seven}) for $Q_1(t)$ into (\ref{eight}),
doing a binomial expansion for the
powers of $Q_0(t)$ and $Q_0(s)$, pairing $\e^{mi\omega t}$ and
$\e^{-mi\omega t}$ terms, and ensemble-averaging gives us
\begin{eqnarray}
\ensemble{\tau_2(t)} =&& {1\over{m\omega}} \intot
  \biggl\{ \sum_{k=0}^N \sum_{i,j=k}^N \sin[(2k+1)\omega(t-s)]
   C_{ijk}(t,s) \nonumber\\
&&\ \ - \sum_{k=1}^N \sum_{i,j=k}^N \sin[(2k-1)\omega(t-s)]
   C_{ijk}(t,s) \nonumber\\
&&\ \ + \sum_{k=1}^N \sum_{i,j=k}^N \sin[2k\omega(t-s)]
   D_{ijk}(t,s) \nonumber\\
&&\ \ - \sum_{k=1}^N \sum_{i,j=k}^N \sin[(2k-2)\omega(t-s)]
   D_{ijk}(t,s) \biggr\}\ ds,
\end{eqnarray}
where
\begin{mathletters}
\begin{equation}
C_{ijk}(t,s) = a^\prime_i(x(t)) a_j(x(s)) (i+j)! (2i+1) (2j+1)
  {{2i}\choose{i-k}} {{2j}\choose{j-k}}
  {\left({{kT}\over{2m\omega^2}}\right)}^{i+j},
\end{equation}
\begin{equation}
D_{ijk}(t,s) = b^\prime_i(x(t)) b_j(x(s)) (i+j-1)! 4ij
  {{2i-1}\choose{i-k}} {{2j-1}\choose{j-k}}
  {\left({{kT}\over{2m\omega^2}}\right)}^{i+j-1}.
\end{equation}
\end{mathletters}
We can collect together and combine those terms with the same sine
factor, to get
\begin{eqnarray}
\ensemble{\tau_2(t)} = && {1\over{m\omega}} \intot
  \biggl\{ \sum_{k=0}^N \sum_{i,j=k}^N \sin[(2k+1)\omega(t-s)]
   E_{ijk}(t,s) \nonumber\\
&&\ \ + \sum_{k=0}^N \sum_{i,j=k}^N \sin[2k\omega(t-s)]
   F_{ijk}(t,s) \biggr\}\ ds,
\end{eqnarray}
where
\begin{mathletters}
\begin{equation}
E_{ijk}(t,s) = a^\prime_i(x(t)) a_j(x(s)) (i+j)!
  {\left({{kT}\over{2m\omega^2}}\right)}^{i+j} (2k+1)(i+j+1)
  {{2i+1}\choose{i-k}} {{2j+1}\choose{j-k}},
\end{equation}
\begin{equation}
F_{ijk}(t,s) = b^\prime_i(x(t)) b_j(x(s)) (i+j-1)!
  {\left({{kT}\over{2m\omega^2}}\right)}^{i+j-1} 2k(i+j)
  {{2i}\choose{i-k}} {{2j}\choose{j-k}}.
\end{equation}
\end{mathletters}

We are also interested in the correlation function $\ensemble{F(t)F(s)}$, where
$F(t)$ is the force due to the interaction with the reservoir.  To
second order this is
\begin{eqnarray}
\ensemble{F(t)F(s)} = && \epsilon^2\biggl\{ \sum_{k=0}^N \sum_{i,j=k}^N
   2\cos[(2k+1)\omega(t-s)] G_{ijk}(t,s) \nonumber\\
&& \ \ + \sum_{k=1}^N \sum_{i,j=k}^N 2\cos[2k\omega(t-s)]
   H_{ijk}(t,s) \nonumber\\
&& \ \ + \sum_{i,j=1}^N H_{ij0}(t,s) \biggr\}
   + O(\epsilon^3),
\end{eqnarray}
\begin{mathletters}
\begin{equation}
G_{ijk}(t,s) = a_i^\prime(x(t)) a_j^\prime(x(s)) (i+j+1)!
  {\left({{kT}\over{2m\omega^2}}\right)}^{i+j+1}
  {{2i+1}\choose{i-k}} {{2j+1}\choose{j-k}},
\end{equation}
\begin{equation}
H_{ijk}(t,s) = b_i^\prime(x(t)) b_j^\prime(x(s)) (i+j)!
  {\left({{kT}\over{2m\omega^2}}\right)}^{i+j}
  {{2i}\choose{i-k}} {{2j}\choose{j-k}}.
\end{equation}
\end{mathletters}
We can subtract off the average values to get
\begin{equation}
\ensemble{F(t),F(s)} = \ensemble{F(t)F(s)}
  - \ensemble{F(t)}\ensemble{F(s)},
\end{equation}
where $\ensemble{F(t)}$ is the first order ensemble averaged force
from (\ref{nine}).

We can compare this result to that obtained from our quantum mechanical
procedure.  Suppose that the reservoir begins in a definite state
$\ket{n}$.  Then the influence functional is given by (\ref{five}),
\begin{equation}
F_n[X(t),\xi(t)] = 1 + \epsilon\alpha_{n1}[X(t),\xi(t)]
  + \epsilon^2\alpha_{n2}[X(t),\xi(t)] + \cdots,
\end{equation}
and in the thermal case by (\ref{six}),
\begin{equation}
F[X(t),\xi(t)] = 1 + \epsilon\alpha_1[X(t),\xi(t)] +
  \epsilon^2\alpha_2[X(t),\xi(t)] +
  \cdots = \sum_n \rho_{nn} F_n[X(t),\xi(t)],
\end{equation}
where
\begin{equation}
\alpha_i[X(t),\xi(t)] = \sum_n \rho_{nn} \alpha_{ni}[X(t),\xi(t)].
\end{equation}
The influence phase is then
\begin{eqnarray}
W[X(t),\xi(t)] =&& -i\hbar \ln F[X(t),\xi(t)] \nonumber\\
  =&& -i\hbar\epsilon\alpha_1[X(t),\xi(t)] \nonumber\\
  &&\ \ - i\hbar\epsilon^2\left(\alpha_2[X(t),\xi(t)] -
  {1\over2}\alpha_1^2[X(t),\xi(t)]\right)
  + \cdots.
\end{eqnarray}
{}From (\ref{five}), then, we see that we must find an expression for
$\bra{m} r^l \ket{n}$.
This will, in general, be a polynomial in $n$, for certain values of $m$,
and zero for the rest.  In comparing to the classical result, we need keep
only the highest power of $n$, since the lower powers will be higher-order
in $\hbar\omega/kT$ as we let $\hbar\rightarrow0$.  This will be
\begin{equation}
\bra{m}r^l\ket{n} = {{l}\choose{k}} {\left({\hbar\over{2m\omega}}\right)}^{l/2}
  n^{l/2} + \cdots, m = n + l - 2k, 2k\le l,
\end{equation}
\begin{equation}
\bra{m}r^l\ket{n} = {{l}\choose{k}} {\left({\hbar\over{2m\omega}}\right)}^{l/2}
  m^{l/2} + \cdots, m = n - l + 2k, 2k\le l,
\end{equation}
and zero otherwise.

We can then use the fact that as $\hbar\rightarrow0$,
\begin{equation}
\sum_n \rho_{nn} n^l \approx l!
  {\left({{kT}\over{\hbar\omega}}\right)}^l.
\end{equation}
Thus, from equation (\ref{five}) we get
\begin{eqnarray}
\alpha_1[X(t),\xi(t)] = && \sum_n \rho_{nn} {i\over\hbar} \intof
   \sum_j b_j^\prime(X(t))\bra{n}r^{2n}\ket{n}\xi(t)\ dt \nonumber\\
= && {i\over\hbar} \intof \sum_j b_j^\prime(X(t)) \xi(t)
      {{2j!}\over{j!}} {\left({{kT}\over{2m\omega^2}}\right)}^j\ dt,
\end{eqnarray}
which agrees exactly with the first order term in the classical
equation of motion (\ref{nine}).

Similarly, we can calculate the second order term to get
\begin{eqnarray}
\alpha_2[X(t),\xi(t)] = && -{1\over{\hbar^2}} \biggl\{
   \sum_{k=0}^N \sum_{i,j=k}^N \intof \intot
   2\cos[(2k+1)\omega(t-s)] G_{ijk}(t,s)\xi(t)\xi(s)\ ds\ dt \nonumber\\
&& \ \ + \sum_{k=1}^N \sum_{i,j=k}^N \intof \intot
   2\cos[2k\omega(t-s)] H_{ijk}(t,s)\xi(t)\xi(s)\ ds\ dt \nonumber\\
&& \ \ + i\sum_{k=1}^N \sum_{i,j=k}^N \intof \intot
   \sin[(2k+1)\omega(t-s)] E_{ijk}(t,s)\xi(t)\ ds\ dt \nonumber\\
&& \ \ + i\sum_{k=1}^N \sum_{i,j=k}^N \intof \intot
   \sin[2k\omega(t-s)] F_{ijk}(t,s)\xi(t)\ ds\ dt \nonumber\\
&& \ \ + \sum_{i,j=1}^N \intof \intot
   H_{ij0}(t,s)\xi(t)\xi(s)\ ds\ dt\biggr\}.
\end{eqnarray}
Here we've used the same definitions of $E_{ijk}$, etc., where the classical
system variable $x$ has become the quantum variable $X$.

We can clearly see from this the exact correspondence with the classical
equation of motion, at least to second order in $\epsilon$.  The real part
of $W[X(t),\xi(t)]$ is just an integral of the classical retarded force,
just as in the linear case, and the imaginary part consists of a double
integral
\begin{equation}
\intof \intot \biggl[ \langle F(t)F(s) \rangle - \langle F(t) \rangle
  \langle F(s) \rangle \biggr]\xi(t)\xi(s)\ ds\ dt;
\end{equation}
note that the $-\langle F(t) \rangle\langle F(s)\rangle$ comes from
subtracting $\alpha_1^2/2$ from the second order term.  Again, we note the
non-negativity of this imaginary part; the presence of noise both makes
the behavior unpredictable, and causes different trajectories to decohere.
So we see that in perturbation theory, the nonlinear problem has exactly the
same classical correspondence as the linear problem.

\section{\bf More general cases}

Though the above discussion is fairly general, it leaves unexamined the far
broader range of possible strong, nonlinear interactions, as well as
the possibilities of non-oscillator reservoirs.  This is, of course, a
product of computational convenience, as it is very difficult to get
analytical answers in other cases.  Are there any arguments that can be
made for more general systems?

In any case where the action can be decomposed
\begin{equation}
S[x(t),Q(t)] = \Ss[x(t)] + \Sr[Q(t)] + \Si[x(t),Q(t)],
\end{equation}
it is possible formally to write the decoherence functional in the form
\begin{equation}
D[x(t),\xp(t)] = \exp {i\over\hbar}
  \biggl\{\Ss[x(t)] - \Ss[\xp(t)] + W[x(t),\xp(t)]\biggr\}.
\end{equation}
If we restrict ourselves, for the moment, to systems in a factorizable
pure state,
\begin{equation}
\rho(\xp,\Qp; x,Q) = \Psi^\ast(\xp)\Psi(x) \Phi^\ast(\Qp)\Phi(Q),
\end{equation}
then this influence phase is defined simply by (2.15)
\begin{eqnarray}
\exp\left\{i W[\xp(t),x(t)]/\hbar\right\}
  = &&\int\delQp\int\delQ\delQf
  \exp {i\over\hbar} \biggl\{ \Sr[\Qp(t)] -
  \Sr[Q(t)] \nonumber\\
  &&\ \ + \Si[\xp(t),\Qp(t)] -
  \Si[x(t),Q(t)] \biggr\}
  \Phi^\ast(\Qp_0)\Phi(Q_0).
\end{eqnarray}

By bringing the integral over the final condition $Q_f,\Qp_f$ to the front,
we can re-write this as a product of two path integrals:
\begin{eqnarray}
\exp\left\{i W[\xp(t),x(t)]/\hbar \right\} = &&
  \int\int dQ_f d\Qp_f \delQf \nonumber\\
  &&\times \Biggl[ \int\delQp \exp {i\over\hbar} \biggl\{ \Sr[\Qp(t)]
  + \Si[\xp(t),\Qp(t)] \biggr\} \Phi^\ast(\Qp_0) \Biggr] \nonumber\\
  &&\times \Biggl[ \int\delQ \exp {i\over\hbar} \biggl\{ - \Sr[Q(t)]
  - \Si[x(t),Q(t)] \biggr\} \Phi(Q_0) \Biggr] \nonumber\\
  = && \int\int dQ_f d\Qp_f \delQf \Phi_{\xp(t)}^\ast(\Qp_f)
    \Phi_{x(t)}(Q_f) \nonumber\\
  = && \int \Phi_{\xp(t)}^\ast(Q_f) \Phi_{x(t)}(Q_f) dQ_f
  = \braket{\Phi_{\xp(t)}}{\Phi_{x(t)}},
\end{eqnarray}
where $\ket{\Phi_{x(t)}}$ and $\ket{\Phi_{\xp(t)}}$ are the states that
$\ket{\Phi}$ will evolve into under the influence of the interaction,
given the trajectories $x(t)$ and $\xp(t)$, respectively.

Clearly, $\braket{\Phi_{\xp(t)}}{\Phi_{x(t)}} \le 1$, which implies equally
clearly that $\im W[\xp(t),x(t)] \ge 0$.  So the non-negativity
that we saw in the cases I and II above is generally true.  This is also
clearly the case for mixed states, since we can represent any mixed state
as
\begin{equation}
\rho(\xp,\Qp; x,Q) = \sum_i p_i \Psi_i^\ast(\xp)\Psi_i(x)
  \Phi_i^\ast(\Qp)\Phi_i(Q),
\end{equation}
where
\begin{equation}
\sum_i p_i = 1,\  p_i \ge 0;
\end{equation}
so if the $F_i[\xp(t),x(t)] < 1$, then clearly
\begin{equation}
\exp\left\{i W[\xp(t),x(t)]/\hbar\right\} = \sum_i p_i F_i[\xp(t),x(t)] \le 1
\end{equation}
and $\im W[\xp(t),x(t)] \ge 0$ still holds.  Also, $\im W[\xp(t),x(t)] = 0$
for $\xp(t) = x(t)$.  Thus, without assuming
{\it anything} about the interaction {\it or} the reservoir, we see that
there will be a maximum at $\xi(t) = 0$, and that the off-diagonal
$\xi(t) \ne 0$ terms will tend to be suppressed.  This is not surprising, as
one expects almost any sort of interaction with neglected degrees of freedom
to result in the loss of phase coherence.  However, it does show how these
highly simplified models might actually demonstrate behavior important to
the rise of classical physics from quantum mechanics in physical systems.

For example, in considering quantum gravity, decoherence might arise from
neglected gravitational degrees of freedom.  The usual semi-classical treatment
of quantum gravity, which omits the ``back action'' of mass-energy on the
curvature of spacetime, cannot exhibit this effect.  The weakness of the
gravitational interaction would in general make it less important in causing
decoherence than stronger forces, such as electromagnetism; but it might well
become important in quantum cosmology.

There are, of course, still questions.  All that has been demonstrated is
the non-negativity of $\im W[\xp(t),x(t)]$.  Can there not be zeroes for
some choices of $\xi(t) \ne 0$?  And {\it how strongly}, in general, are
the off-diagonal terms suppressed?

There can certainly be zeroes for nonzero $\xi(t)$ in some cases.  Indeed,
if we consider the form of $\im W[\xp(t),x(t)]$ for the linear case
\begin{equation}
\im W[\xp(t),x(t)] \sim \intof\intof \xi(t)\xi(s)\cos[\omega(t-s)]\ ds\ dt
\end{equation}
(for a one-oscillator ``reservoir'' of frequency $\omega$), there are an
infinite number of choices of $\xi(t)$ which make this zero.  Thus, one
cannot call this system truly decoherent.  However, as the number of
oscillator frequencies is increased, the number of
possible choices of $\xi(t)$ is
further and further restricted, so that as the reservoir becomes infinite
only $\xi(t) = 0$ remains.  One would expect similar behavior in the
more general case.  While it is certainly possible to construct cases where
$\im W[\xp(t),x(t)]$ has many zeroes even for a very large reservoir, in
practice one expects $\im W[\xp(t),x(t)] > 0$ for $x(t) \ne \xp(t)$, as
the degrees of freedom of the reservoir are increased.

Similarly, the strength with which off-diagonal terms will be suppressed
depends on the details of the system.  However, one would expect that
$\ket{\Phi_{x(t)}}$ and $\ket{\Phi_{\xp(t)}}$ differ more in the case of strong
interactions than small, and hence that $\braket{\Phi_{\xp(t)}}{\Phi_{x(t)}}$
would be more strongly suppressed, in general.

\section{\bf Conclusions}

It is clear that it is possible to define a ``classical'' equation of motion
directly from the underlying quantum theory, and that, at least in many
cases, this corresponds closely to the equation obtained from the classical
theory.  While correspondences of this sort have often been demonstrated
in the past, never before has there been a rigorous, {\it a priori} technique
for deriving them.

Using the formalism of Gell-Mann and Hartle, we can now see classical
physics as, very simply, a limit of the underlying quantum theory; and we
can systematically determine, at least in principle, the deviations from
strict classical equations due to quantum effects.  Using the decoherence
functional as a criterion for determining if an effect is experimentally
observable, we can once and for all avoid the problem of collapsing the
wave function; there is no longer any need for an independent
``classical realm'' of measurement.

\acknowledgements

I would like to acknowledge the guidance and encouragement of Murray Gell-Mann,
to whom I owe any success in this area, and James Hartle for his lucid
insight and patient corrections;
also to Seth Lloyd, who patiently listened to my numerous complaints,
questions, and ideas, and offered many helpful suggestions.

After the completion of this research, I learned that Bei Lok Hu, Juan Pablo
Paz, and Yuhong Zhang had studied a very similar class of nonlinear brownian
systems more or less simultaneously with me \cite{HuPazZhang}.
While their study is from a
considerably different point of view, with very different goals,
being chiefly concerned with deriving master equations and
treating the thermodynamics of these generalized systems, their
results overlap mine to a certain extent.

\end{document}